\documentclass[conference]{IEEEtran}
\IEEEoverridecommandlockouts
\usepackage{amsmath, amssymb, graphicx, xcolor, multirow, multicol, comment, subcaption, algorithm2e, url, float, etoolbox, adjustbox, pgf, soul, geometry, colortbl, booktabs}
\definecolor{salmonlight5}{RGB}{255,228,225}   
\definecolor{salmonlight10}{RGB}{250,210,200}    
\definecolor{salmonlight15}{RGB}{245,190,175}    
\definecolor{salmonlight20}{RGB}{240,170,150}    
\definecolor{salmonlight25}{RGB}{235,150,125}    

\definecolor{salmonlight5}{RGB}{255,228,225}   
\definecolor{salmonlight10}{RGB}{250,210,200}    
\definecolor{salmonlight15}{RGB}{245,190,175}    
\definecolor{salmonlight30}{RGB}{235,150,125}    

\definecolor{darkconcat5}{RGB}{245,218,213}    
\definecolor{darkconcat10}{RGB}{240,200,185}     
\definecolor{darkconcat15}{RGB}{235,180,165}     
\definecolor{darkconcat20}{RGB}{230,160,145}     

\definecolor{darkreno10}{RGB}{240,200,185}       
\definecolor{darkreno15}{RGB}{235,180,165}       
\definecolor{darkreno20}{RGB}{230,160,145}       
\definecolor{darkreno25}{RGB}{225,140,125}       

\definecolor{concat5}{RGB}{255,228,225}    
\definecolor{concat10}{RGB}{250,210,200}     
\definecolor{concat15}{RGB}{245,190,175}     
\definecolor{concat20}{RGB}{240,170,150}     

\definecolor{reno10}{RGB}{250,210,200}       
\definecolor{reno15}{RGB}{245,190,175}       
\definecolor{reno20}{RGB}{240,170,150}       
\definecolor{reno25}{RGB}{235,150,125}       

\definecolor{darkconcat5}{RGB}{245,218,213}    
\definecolor{darkconcat10}{RGB}{240,200,185}     
\definecolor{darkconcat15}{RGB}{235,180,165}     
\definecolor{darkconcat20}{RGB}{230,160,145}     
\definecolor{darkconcat25}{RGB}{225,140,125}     

\definecolor{darkreno10}{RGB}{250,220,210}       
\definecolor{darkreno15}{RGB}{245,200,185}       
\definecolor{darkreno20}{RGB}{240,180,165}       
\definecolor{darkreno25}{RGB}{235,160,145}     
\definecolor{reno5}{RGB}{255,240,240}  

\title{ARE MAMBA-BASED AUDIO FOUNDATION MODELS THE BEST FIT FOR NON-VERBAL EMOTION RECOGNITION?}

\author{
    \IEEEauthorblockN{
        Mohd Mujtaba Akhtar\IEEEauthorrefmark{1}\textsuperscript{\dag},
        Orchid Chetia Phukan\IEEEauthorrefmark{2}\textsuperscript{\dag},
Girish\IEEEauthorrefmark{2}\IEEEauthorrefmark{3}\textsuperscript{\dag}       
        Swarup Ranjan Behera\IEEEauthorrefmark{4}\\
        Ananda Chandra Nayak\IEEEauthorrefmark{5},
        Sanjib Kumar Nayak\IEEEauthorrefmark{6},
        Arun Balaji Buduru\IEEEauthorrefmark{1},
        Rajesh Sharma\IEEEauthorrefmark{7}\IEEEauthorrefmark{8}
    }
    \IEEEauthorblockA{
    \IEEEauthorrefmark{1}\textit{V.B.S.P.U, India},
        \IEEEauthorrefmark{2}\textit{IIIT-Delhi, India}, 
        \IEEEauthorrefmark{3}\textit{UPES, India},
        \IEEEauthorrefmark{4}\textit{Independent Researcher, India}\\
        \IEEEauthorrefmark{5}\textit{KAC, India}, 
        \IEEEauthorrefmark{6}\textit{VSSUT, India},
        \IEEEauthorrefmark{7}\textit{University of Tartu, Estonia},
         \IEEEauthorrefmark{8}\textit{Plaksha University,India}\\
        \texttt{\textcolor{blue}{Correspondence:}mmakhtar.research@gmail.com,orchidp@iiitd.ac.in}
    }
    \thanks{\textsuperscript{\dag} Contributed equally as a first authors}
}

\begin{document}

\maketitle

\begin{abstract}

In this work, we focus on non-verbal vocal sounds emotion recognition (NVER). We investigate mamba-based audio foundation models (MAFMs) for the first time for NVER and hypothesize that MAFMs will outperform attention-based audio foundation models (AAFMs) for NVER by leveraging its state-space modeling to capture intrinsic emotional structures more effectively. Unlike AAFMs, which may amplify irrelevant patterns due to their attention mechanisms, MAFMs will extract more stable and context-aware representations, enabling better differentiation of subtle non-verbal emotional cues. Our experiments with state-of-the-art (SOTA) AAFMs and MAFMs validates our hypothesis. Further, motivated from related research such as speech emotion recognition, synthetic speech detection, where fusion of foundation models (FMs) have showed improved performance, we also explore fusion of FMs for NVER. To this end, we propose, \texttt{\textbf{RENO}}, that uses renyi-divergence as a novel loss function for effective alignment of the FMs. It also makes use of self-attention for better intra-representation interaction of the FMs. With \textbf{\texttt{RENO}}, through the heterogeneous fusion of MAFMs and AAFMs, we show the topmost performance in comparison to individual FMs, its fusion and also setting SOTA in comparison to previous SOTA work. 
\end{abstract}

\section{INTRODUCTION \& RELATED WORK}

Emotions play a fundamental role in human communication, shaping how we express ourselves and connect with others. They influence not only verbal interactions but also non-verbal cues, such as facial expressions, gestures, physiological signals, and vocalizations, which are crucial in conveying underlying feelings and intentions. However, non-verbal vocalizations offer a unique and often underexplored perspective. Sounds like laughter, cries, and sighs convey a broad spectrum of emotions that play a crucial role in communication, enhancing human interactions in daily life. Recognizing emotions from these non-verbal vocal cues has diverse applications in areas such as healthcare, human-computer interaction, customer service, and security, where understanding emotional context is vital for decision-making and improving user experience. Unlike speech emotion recognition (SER), which relies on language, these vocalizations bypass the need for verbal content and instead communicate emotions directly through their acoustic features. However, SER has been extensively studied in comparison to non-verbal vocal sounds emotion recognition (NVER). \par
Initial works on SER focused on the usage of handcrafted spectral features such as MFCC with classical ML algorithms such as SVM, HMM, GMM, decision tree \cite{lin2005speech, kishore2013emotion, liu2018speech}. Researchers have also leveraged handcrafted features with neural network-based approaches such as RNN, CNN, LSTM, Transformer \cite{wang2020speech, pawar2021convolution, lian2021ctnet}. However, with the advent of foundation models (FMs) in recent years, SER research has significantly shifted towards the use of FMs \cite{pepino21_interspeech, phukan2023transforming, diatlova2024adapting}. These models, pre-trained on large datasets, offer superior performance and the ability to capture complex acoustic patterns, reducing the need for manually engineered features and enabling more robust emotion recognition in diverse contexts. These FMs are either attention-based audio FMs (AAFMs) and mamba-based audio FMs (MAFMs). MAFMs built on state-space models (SSMs), which offer a computationally efficient alternative to traditional attention-based architectures. Unlike attention mechanisms that dynamically assign weights to input features, SSMs model sequences through structured recurrence, enabling long-range dependency capture while maintaining scalability. As such benefits, researchers have explored building audio FMs with mamba-based modeling architectures and they have shown superior or comparative performance in SER in comparison to AAFMs \cite{yadav24_interspeech}.\par
Despite much advancement in SER, research into NVER haven't seen much limelight except few prolific works \cite{hsu2021speech, xin2024jvnv} despite carrying sufficient potential for emotion recognition. Audio FMs such as Wav2vec2, Whisper have also shown its efficacy for NVER \cite{tzirakis2023large}. However, the Audio FMs used in previous research are mostly AAFMs and previous works haven't investigated MAFMs for NVER and this leaves a gap towards understanding the potential of MAFMs for NVER. In this work, we focus on NVER and explore MAFMs for NVER, to the best of knowledge. \textit{We hypothesize that MAFMs will outperform attention-based AAFMs in NVER by leveraging their state-space modeling capabilities for better capture of intrinsic emotional structures. In contrast to AAFMs, which may amplify irrelevant patterns due to their attention mechanisms, MAFMs offer more stable and contextually aware representations, facilitating the differentiation of subtle non-verbal emotional cues}. Our experiments, comparing state-of-the-art (SOTA) AAFMs and MAFMs, confirm the validity of our hypothesis. Furthermore, drawing inspiration from related fields such as speech emotion recognition \cite{10317508} and synthetic speech detection \cite{chetia-phukan-etal-2024-heterogeneity}, where the fusion of FMs has led to performance improvements, we investigate the fusion of FMs for NVER. To achieve this, we propose \texttt{\textbf{RENO}}, (\texttt{\textbf{REN}}yi Attenti\texttt{\textbf{O}}n Network), a novel framework to to effectively align the FMs. \texttt{\textbf{RENO}} incorporates self-attention mechanisms to enhance intra-representation interactions across the FMs representational spacne and utilizes Renyi-divergence as a novel loss function for inter-FM interaction. Through \texttt{\textbf{RENO}} with the heterogeneous fusion of MAFMs and AAFMs, we demonstrate superior performance outperforming both individual MAFMs, AAFMs, baseline fusion methods as well as homogeneous fusion of AAFMs, thus setting a new SOTA for NVER. 

\noindent \textbf{To summarize, the main contributions of this study are as follows:}
\begin{itemize}
    \item For investigating the effectiveness of MAFMs for NVER, we present the first comprehensive comparative study of MAFMs and AAFMs. Our experiments results shows that MAFMs outperforms its attention-based counterparts.
    \item We propose, \textbf{\texttt{RENO}}, a novel framework that leverages self-attention for intra-FM interaction followed by the usage of Renyi Divergence loss for inter-FMs alignment. \textbf{\texttt{RENO}} with the heterogenous fusion of MAFMs and AAFMs shows the topmost performance in comparison to individual FMs, baseline fusion techniques, homogeneous fusion of AAFMs, and thus achieving SOTA across benchmark NVER datasets such as ASVP-ESD, JNV, and VIVAE. 
\end{itemize}
\noindent The code and models developed in this study can be accessed at \footnote{\url{https://github.com/Helix-IIIT-Delhi/RENO-Non-Verbal}}.

\section{Foundation Models}
In this section, we provide an overview of the FMs used in our study.

\noindent \textbf{Audio-MAMBA\footnote{\url{https://github.com/SarthakYadav/audio-mamba-official?tab=readme-ov-file}}} \cite{yadav24_interspeech}: It is a selective SSM designed to learn general-purpose audio representations through self-supervised learning. It extracts information from randomly masked spectrogram patches. Pre-trained on the AudioSet dataset, it consistently shows comparable and sometimes better performance than AAFMs across various tasks, including SER. In our study, we use three versions of Audio-MAMBA: tiny (4.8M parameters), small (17.9M parameters), and base (69.3M parameters).  

\noindent \textbf{WavLM}\footnote{\url{https://huggingface.co/microsoft/wavlm-base}} \cite{chen2022wavlm}: It is a SOTA AAFM, ranked highly on the SUPERB benchmark. It uses masked speech modeling with denoising objectives during its pre-training. We employ the base version, which consists of 94.70M parameters and is trained on 960 hours of English speech from the LibriSpeech dataset.  

\noindent \textbf{UniSpeech-SAT}\footnote{\url{https://huggingface.co/microsoft/unispeech-sat-base}} \cite{chen2022unispeech}: It is another SOTA AAFM on the SUPERB leaderboard. It follows a self-supervised pre-training approach with speaker-aware multi-task learning. We utilize the base version, which has 94.68M parameters and is trained on 960 hours of English speech from LibriSpeech.  

\noindent \textbf{Wav2vec2}\footnote{\url{https://huggingface.co/facebook/wav2vec2-base}} \cite{baevski2020wav2vec}: It is an AAFM that employs contrastive self-supervised learning to learn speech representations. It masks segments of latent features and optimizes them through contrastive loss. We use its base version, which contains 95.04M parameters and is pre-trained on 960 hours of English speech from the LibriSpeech dataset.  

\noindent \textbf{HuBERT}\footnote{\url{https://huggingface.co/facebook/hubert-base-ls960}} \cite{hsu2021hubert}: It follows a self-supervised learning framework that iteratively refines its representations using k-means clustering while training on a BERT-style masked prediction objective. We employ the base version, which has 94.68M parameters and is trained on 960 hours of English speech from LibriSpeech.

\noindent Resampling to 16 KHz is done for the audio samples before passing it to the FMs. We obtain representations from the last hidden state of the frozen FMs including MAFMs and AAFMs using average pooling. The extracted representations have the following dimensions: 768 for WavLM, UniSpeech-SAT, Wav2Vec 2.0, and HuBERT; 1280 for MMS; and for Audio-MAMBA, 960 for the tiny version, 1920 for the small version, and 3840 for the base version.

\section{Modeling Methodology}
In this section, we discuss the downstream modeling used with the FMs followed by the discussion of the proposed framework, \textbf{\texttt{RENO}} for the fusion of FMs. We utilize FCN (Fully connected network) and CNN as downstream classifiers. The CNN model consists of two convolutional blocks consisting of 1D CNN layers with 32 and 64 filters, respectively with filter size of 3. Max-pooling is applied after each convolutional layer. The extracted features are then flattened and passed through a FCN block with two dense layers of 512 and 128 neurons followed by the output layer with softmax activation function. The output layer outputs probabilities for the emotion classes. The FCN follows the same modeling as the FCN block of the CNN model. CNN models training parameters varies between 0.9 to 1.1M while FCN models parameters between 0.7 to 1M depending on the dimension size of input representations.

\subsection{\textbf{\texttt{RENO}}}

\begin{figure}[!bt]
    \centering
    \includegraphics[width=0.62\linewidth]{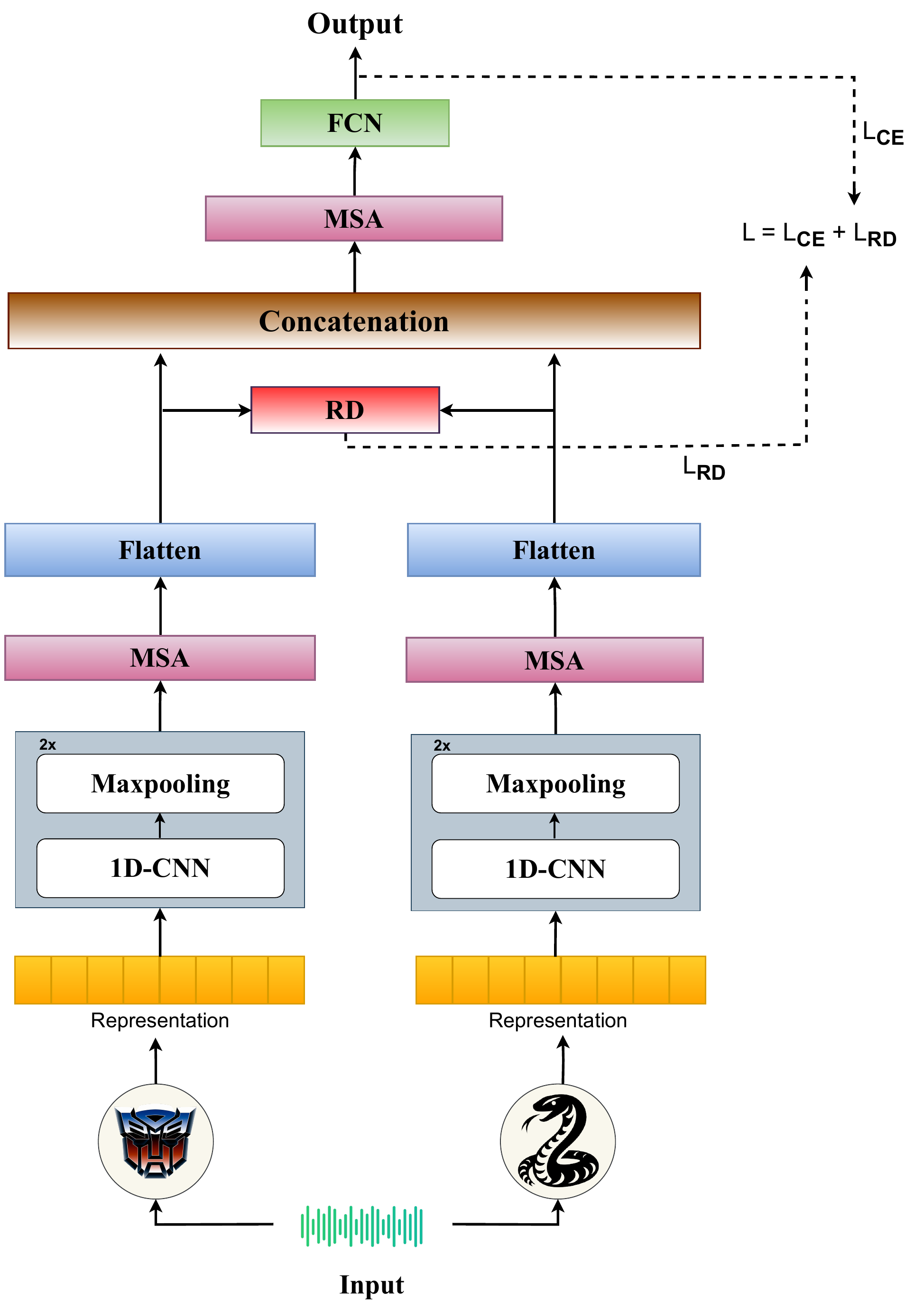}
    \caption{Novel Framework: \textbf{\texttt{RENO}}; MSA stands for Multi-head Self-attention}
    \label{fig:proposed}
\end{figure}

We propose \texttt{\textbf{RENO}}, a novel framework designed to align feature representations from distinct FMs. The architecture is given in Figure \ref{fig:proposed}. \texttt{\textbf{RENO}} leverages self-attention mechanisms to strengthen intra-representation interactions within the feature space of FMs followed by employing Rényi divergence (RD) as a novel loss function to facilitate inter-FM interaction. Detailed walkthrough of the proposed framework is given as follows: The extracted representations from different FMs are first flattened after passing through convolutional block as used with individual FM representations above. We then passed the flattened features through a self-attention mechanism for better intra-representation interaction. Self-attention is calculated as follows:
\begin{equation}
    \text{Attention}(\mathbf{Q}, \mathbf{K}, \mathbf{V}) = \text{softmax} \left( \frac{\mathbf{Q} \mathbf{K}^T}{\sqrt{d_k}} \right) \mathbf{V}
\end{equation}

\noindent  where $d_k$ is the scaling factor. $\mathbf{Q}$, $\mathbf{K}$, $\mathbf{Q}$ stands for query, key, value where $ \mathbf{Q} = \mathbf{W}_Q \mathbf{X}, \quad \mathbf{K} = \mathbf{W}_K \mathbf{X}, \quad \mathbf{V} = \mathbf{W}_V \mathbf{X} $. $\mathbf{X}$ represents the input feature matrix and $\mathbf{W}_Q, \mathbf{W}_K, \mathbf{W}_V$ are learnable weight matrices. Then the features are passed through to RD loss. RD quantifies the difference or dissimilarity between two probability distributions \cite{van2014renyi}. Lower RD higher the similarity. Here in our study, we introduce RD as novel loss function for measuring the divergence between feature distributions of two FMs. Given two feature spaces $e_{a}$ and $e_{b}$ corresponding to two FMs, RD is computed as:

\begin{equation}
    \mathcal{L}_{RD} = \frac{1}{\beta - 1} \log \left( \sum_{j=1}^M (z_{x,j} + \delta)^{\beta} (z_{y,j} + \delta)^{1 - \beta} \right)
\end{equation}

where $M$ is the feature dimension, $\beta > 1$ controls the divergence order, and $\delta$ ensures numerical stability. RD will align the representation space of the FMs to a joint feature space and following this, the aligned features are then concatenated and passed through a self attention block. Self-attention will lead to further refinement of the fused features. Finally, the features are passed through a FCN block with two dense layers with each layer consisting of 512 and 128 neurons. The output layer leverages softmax as activation and outputs probabilities for the emotion classes.
For joint optimization with cross-entropy loss $\mathcal{L}_{CE}$, we integrate RD loss $\mathcal{L}_{RD}$ with the $\mathcal{L}_{CE}$:

\begin{equation}
    \mathcal{L} = \lambda \mathcal{L}_{CE} + (1 - \lambda) \mathcal{L}_{RD}
\end{equation}

\noindent where $\lambda$ is a hyperparameter controlling the trade-off between $\mathcal{L}_{CE}$ and $\mathcal{L}_{RD}$. The training parameters varies between 1.3 to 1.5M depending on the dimensional-size of the input representations. For self-attention blocks for each individual repesentation network, we keep the number of heads as 2 and for the self-attention block after concatenation, we set the number of heads also to 2.

\begin{table}[!bt]
\scriptsize
\setlength{\tabcolsep}{9pt}
\centering
\begin{tabular}{l|ll|ll|ll}
\toprule
\multicolumn{1}{c|}{\textbf{FM}} & \multicolumn{2}{c|}{\textbf{ASVP\_ESD}} & \multicolumn{2}{c|}{\textbf{JNV}} & \multicolumn{2}{c}{\textbf{VIVAE}} \\ 
\midrule
                                  & \textbf{A} \(\uparrow\) & \textbf{F1} \(\uparrow\) & \textbf{A} \(\uparrow\) & \textbf{F1} \(\uparrow\) & \textbf{A} \(\uparrow\) & \textbf{F1} \(\uparrow\) \\
\midrule
\multicolumn{7}{c}{\textbf{FCN}} \\
\cmidrule{1-7}
A (T) & \cellcolor{salmonlight20}66.51 & \cellcolor{salmonlight10}58.71 & \cellcolor{salmonlight10}59.82 & \cellcolor{salmonlight10}58.14 & \cellcolor{salmonlight5}52.21 & \cellcolor{salmonlight5}51.95 \\
A (S) & \cellcolor{salmonlight25}71.15 & \cellcolor{salmonlight20}68.74 & \cellcolor{salmonlight15}61.17 & \cellcolor{salmonlight15}60.84 & \cellcolor{salmonlight10}57.48 & \cellcolor{salmonlight10}56.31 \\
A (B) & \cellcolor{salmonlight25}72.64 & \cellcolor{salmonlight20}69.47 & \cellcolor{salmonlight15}62.28 & \cellcolor{salmonlight15}60.97 & \cellcolor{salmonlight10}58.99 & \cellcolor{salmonlight10}58.57 \\

W     & \cellcolor{salmonlight5}52.69  & \cellcolor{salmonlight5}42.86 & \cellcolor{salmonlight5}57.46 & \cellcolor{salmonlight5}56.21 & \cellcolor{salmonlight5}32.54 & \cellcolor{salmonlight5}31.82 \\
W2    & \cellcolor{salmonlight15}61.18 & \cellcolor{salmonlight5}54.32  & \cellcolor{salmonlight10}56.96 & \cellcolor{salmonlight10}55.14 & \cellcolor{salmonlight5}47.85 & \cellcolor{salmonlight5}46.81 \\
U     & \cellcolor{salmonlight5}54.96  & \cellcolor{salmonlight5}53.21 & \cellcolor{salmonlight10}57.52 & \cellcolor{salmonlight10}56.62 & \cellcolor{salmonlight5}34.28 & \cellcolor{salmonlight5}33.68 \\
H     & \cellcolor{salmonlight10}55.67 & \cellcolor{salmonlight5}54.05  & \cellcolor{salmonlight10}58.41 & \cellcolor{salmonlight10}57.96 & \cellcolor{salmonlight5}42.56 & \cellcolor{salmonlight5}41.94 \\
\midrule
\multicolumn{7}{c}{\textbf{CNN}} \\
\cmidrule{1-7}
A (T) & \cellcolor{salmonlight20}67.59 & \cellcolor{salmonlight20}65.84  & \cellcolor{salmonlight15}62.77 & \cellcolor{salmonlight15}61.14 & \cellcolor{salmonlight5}53.92 & \cellcolor{salmonlight5}53.57 \\
A (S) & \cellcolor{salmonlight25}72.90 & \cellcolor{salmonlight25}70.64  & \cellcolor{salmonlight15}63.10 & \cellcolor{salmonlight15}61.42 & \cellcolor{salmonlight15}60.41 & \cellcolor{salmonlight10}59.93 \\
\textbf{A (B)} & \cellcolor{salmonlight25}\textbf{73.96} & \cellcolor{salmonlight25}\textbf{71.98}  & \cellcolor{salmonlight15}\textbf{64.29} & \cellcolor{salmonlight15}\textbf{63.81} & \cellcolor{salmonlight15}\textbf{62.42} & \cellcolor{salmonlight15}\textbf{61.29} \\
W     & \cellcolor{salmonlight5}53.94  & \cellcolor{salmonlight5}50.84  & \cellcolor{salmonlight10}59.85 & \cellcolor{salmonlight10}58.89 & \cellcolor{salmonlight5}33.18 & \cellcolor{salmonlight5}32.76 \\
W2    & \cellcolor{salmonlight15}62.96 & \cellcolor{salmonlight15}60.84  & \cellcolor{salmonlight10}59.27 & \cellcolor{salmonlight10}57.14 & \cellcolor{salmonlight5}48.39 & \cellcolor{salmonlight5}48.18 \\
U     & \cellcolor{salmonlight10}55.41 & \cellcolor{salmonlight5}54.16  & \cellcolor{salmonlight15}60.71 & \cellcolor{salmonlight10}59.20 & \cellcolor{salmonlight5}36.69 & \cellcolor{salmonlight5}35.91 \\
H     & \cellcolor{salmonlight10}59.43 & \cellcolor{salmonlight10}58.21  & \cellcolor{salmonlight10}59.26 & \cellcolor{salmonlight10}58.35 & \cellcolor{salmonlight5}49.04 & \cellcolor{salmonlight5}48.36 \\
\bottomrule
\end{tabular}
\caption{Evaluation Scores are in \%; A and F1 stands for Accuracy and macro-average F1 score; Audio-mamba (Tiny: A(T), Small: A(S), Base: A(B)), WavLM (W), Wav2vec2 (W2), and Unispeech-SAT (U); Evaluation scores are given in average across five folds; The abbreviations used are kept same for Table \ref{tab:2}}
\label{tab-1}
\end{table}

\begin{table*}[!bt]
\scriptsize
\setlength{\tabcolsep}{8pt}
\centering
\begin{tabular}{l|cc|cc|cc|cc|cc|cc}
\toprule
\multicolumn{1}{c|}{\textbf{}} & \multicolumn{4}{c|}{\textbf{ASVP\_ESD}} & \multicolumn{4}{c|}{\textbf{JNV}} & \multicolumn{4}{c}{\textbf{VIVAE}} \\ 
\midrule
                                    & \multicolumn{2}{c|}{\textbf{Concat}} & \multicolumn{2}{c|}{\textbf{\texttt{RENO}}} & \multicolumn{2}{c|}{\textbf{Concat}} & \multicolumn{2}{c|}{\textbf{\texttt{RENO}}} & \multicolumn{2}{c|}{\textbf{Concat}} & \multicolumn{2}{c}{\textbf{\texttt{RENO}}} \\ 
\cmidrule(lr){2-5} \cmidrule(lr){6-9} \cmidrule(lr){10-13}
\multicolumn{1}{l|}{\textbf{Combinations}} 
                                    & \textbf{A} \(\uparrow\) & \textbf{F1} \(\uparrow\) & \textbf{A} \(\uparrow\)& \textbf{F1} \(\uparrow\) & \textbf{A} \(\uparrow\)& \textbf{F1} \(\uparrow\) & \textbf{A} \(\uparrow\)& \textbf{F1} \(\uparrow\) & \textbf{A} \(\uparrow\)& \textbf{F1} \(\uparrow\) & \textbf{A} \(\uparrow\)& \textbf{F1} \(\uparrow\) \\ 
\midrule
\multicolumn{13}{c}{\textbf{AAFMs + MAFMs}} \\
\midrule
A (T)+W  & 
  \cellcolor{darkconcat15}70.32 & \cellcolor{darkconcat15}68.91 & 
  \cellcolor{darkreno20}75.42 & \cellcolor{darkreno20}74.23 & 
  \cellcolor{darkconcat15}66.14 & \cellcolor{darkconcat15}65.43 & 
  \cellcolor{darkreno20}71.45 & \cellcolor{darkreno20}70.32 & 
  \cellcolor{darkconcat5}54.29  & \cellcolor{darkconcat5}53.21 & 
  \cellcolor{darkreno10}61.65 & \cellcolor{darkreno10}60.45 \\
A (T)+W2 & 
  \cellcolor{darkconcat10}68.54 & \cellcolor{darkconcat10}67.74 & 
  \cellcolor{darkreno20}74.23 & \cellcolor{darkreno20}73.56 & 
  \cellcolor{darkconcat10}59.42 & \cellcolor{darkconcat10}58.92 & 
  \cellcolor{darkreno10}64.54 & \cellcolor{darkreno10}63.76 & 
  \cellcolor{darkconcat10}56.61 & \cellcolor{darkconcat10}55.03 & 
  \cellcolor{darkreno10}62.34 & \cellcolor{darkreno10}61.32 \\
A (T)+U  & 
  \cellcolor{darkconcat15}71.89 & \cellcolor{darkconcat15}70.45 & 
  \cellcolor{darkreno20}77.34 & \cellcolor{darkreno20}76.12 & 
  \cellcolor{darkconcat15}65.45 & \cellcolor{darkconcat15}64.41 & 
  \cellcolor{darkreno20}72.45 & \cellcolor{darkreno20}71.87 & 
  \cellcolor{darkconcat10}56.87 & \cellcolor{darkconcat10}55.91 & 
  \cellcolor{darkreno10}63.76 & \cellcolor{darkreno10}62.43 \\
A (T)+H  & 
  \cellcolor{darkconcat15}74.41 & \cellcolor{darkconcat15}73.71 & 
  \cellcolor{darkreno25}78.71 & \cellcolor{darkreno25}77.89 & 
  \cellcolor{darkconcat25}\textbf{76.56} & \cellcolor{darkconcat25}\textbf{75.43} & 
  \cellcolor{darkreno25}\textbf{79.09} & \cellcolor{darkreno25}\textbf{78.41} & 
  \cellcolor{darkconcat10}56.53 & \cellcolor{darkconcat10}55.72 & 
  \cellcolor{darkreno10}61.45 & \cellcolor{darkreno10}61.03 \\
A (S)+W  & 
  \cellcolor{darkconcat15}75.08 & \cellcolor{darkconcat15}74.78 & 
  \cellcolor{darkreno25}81.65 & \cellcolor{darkreno25}80.23 & 
  \cellcolor{darkconcat15}66.31 & \cellcolor{darkconcat15}65.79 & 
  \cellcolor{darkreno20}71.67 & \cellcolor{darkreno20}70.23 & 
  \cellcolor{darkconcat15}64.43 & \cellcolor{darkconcat15}62.99 & 
  \cellcolor{darkreno15}73.65 & \cellcolor{darkreno15}72.49 \\
A (S)+W2 & 
  \cellcolor{darkconcat15}73.06 & \cellcolor{darkconcat15}72.32 & 
  \cellcolor{darkreno20}79.34 & \cellcolor{darkreno20}78.31 & 
  \cellcolor{darkconcat15}60.56 & \cellcolor{darkconcat10}59.23 & 
  \cellcolor{darkreno10}67.29 & \cellcolor{darkreno10}66.31 & 
  \cellcolor{darkconcat15}64.21 & \cellcolor{darkconcat15}63.78 & 
  \cellcolor{darkreno25}70.31 & \cellcolor{darkreno25}69.63 \\
A (S)+U  & 
  \cellcolor{darkconcat15}73.67 & \cellcolor{darkconcat15}72.65 & 
  \cellcolor{darkreno25}82.44 & \cellcolor{darkreno25}81.34 & 
  \cellcolor{darkconcat15}63.98 & \cellcolor{darkconcat15}62.54 & 
  \cellcolor{darkreno10}68.43 & \cellcolor{darkreno10}67.60 & 
  \cellcolor{darkconcat15}64.42 & \cellcolor{darkconcat15}63.02 & 
  \cellcolor{darkreno15}71.34 & \cellcolor{darkreno15}70.56 \\
A (S)+H  & 
  \cellcolor{darkconcat15}75.23 & \cellcolor{darkconcat15}74.62 & 
  \cellcolor{darkreno20}76.20 & \cellcolor{darkreno20}75.45 & 
  \cellcolor{darkconcat25}74.37 & \cellcolor{darkconcat25}73.11 & 
  \cellcolor{darkreno25}77.56 & \cellcolor{darkreno25}76.83 & 
  \cellcolor{darkconcat10}57.83 & \cellcolor{darkconcat10}56.72 & 
  \cellcolor{darkreno20}62.47 & \cellcolor{darkreno20}61.43 \\
A (B)+W  & 
  \cellcolor{darkconcat15}75.21 & \cellcolor{darkconcat15}74.47 & 
  \cellcolor{darkreno25}83.56 & \cellcolor{darkreno25}82.54 & 
  \cellcolor{darkconcat15}66.74 & \cellcolor{darkconcat15}65.27 & 
  \cellcolor{darkreno25}72.56 & \cellcolor{darkreno25}71.09 & 
  \cellcolor{darkconcat15}63.70 & \cellcolor{darkconcat15}62.65 & 
  \cellcolor{darkreno25}72.76 & \cellcolor{darkreno25}71.65 \\
A (B)+W2 & 
  \cellcolor{darkconcat15}75.54 & \cellcolor{darkconcat15}74.78 & 
  \cellcolor{darkreno25}82.65 & \cellcolor{darkreno25}81.34 & 
  \cellcolor{darkconcat15}67.62 & \cellcolor{darkconcat15}66.09 & 
  \cellcolor{darkreno25}72.56 & \cellcolor{darkreno25}71.50 & 
  \cellcolor{darkconcat15}61.03 & \cellcolor{darkconcat15}60.34 & 
  \cellcolor{darkreno25}68.93 & \cellcolor{darkreno25}67.32 \\
A (B)+U  & 
  \cellcolor{darkconcat15}\textbf{75.67} & \cellcolor{darkconcat15}\textbf{74.84} & 
  \cellcolor{darkreno25}\textbf{83.56} & \cellcolor{darkreno25}\textbf{82.51} & 
  \cellcolor{darkconcat15}66.72 & \cellcolor{darkconcat15}65.87 & 
  \cellcolor{darkreno25}69.70 & \cellcolor{darkreno25}69.04 & 
  \cellcolor{darkconcat15}61.94 & \cellcolor{darkconcat15}60.13 & 
  \cellcolor{darkreno25}68.53 & \cellcolor{darkreno25}67.08 \\
A (B)+H  & 
  \cellcolor{darkconcat15}75.09 & \cellcolor{darkconcat15}74.21 & 
  \cellcolor{darkreno20}77.98 & \cellcolor{darkreno20}76.48 & 
  \cellcolor{darkconcat15}67.88 & \cellcolor{darkconcat15}66.60 & 
  \cellcolor{darkreno20}71.90 & \cellcolor{darkreno20}71.62 & 
  \cellcolor{darkconcat15}\textbf{67.69} & \cellcolor{darkconcat15}\textbf{66.51} & 
  \cellcolor{darkreno25}\textbf{72.51} & \cellcolor{darkreno25}\textbf{71.82} \\
\midrule
\multicolumn{13}{c}{\textbf{AAFMs + AAFMs}} \\
\midrule
W+W2  & 
  \cellcolor{concat15}63.93 & \cellcolor{concat15}62.32 & 
  \cellcolor{reno20}71.45 & \cellcolor{reno20}70.59 & 
  \cellcolor{concat15}62.01 & \cellcolor{concat15}61.97 & 
  \cellcolor{reno15}69.39 & \cellcolor{reno15}68.51 & 
  \cellcolor{concat5}48.39  & \cellcolor{concat5}47.93 & 
  \cellcolor{reno10}59.31 & \cellcolor{reno10}58.47 \\
W+U   & 
  \cellcolor{concat10}56.73 & \cellcolor{concat10}55.62 & 
  \cellcolor{reno20}67.31 & \cellcolor{reno20}66.10 & 
  \cellcolor{concat15}64.34 & \cellcolor{concat15}63.97 & 
  \cellcolor{reno20}73.41 & \cellcolor{reno20}72.89 & 
  \cellcolor{concat5}38.56  & \cellcolor{concat5}37.54 & 
  \cellcolor{reno10}47.99  & \cellcolor{reno10}46.81 \\
W2+U  & 
  \cellcolor{concat15}65.32 & \cellcolor{concat15}64.75 & 
  \cellcolor{reno20}74.78 & \cellcolor{reno20}73.01 & 
  \cellcolor{concat15}64.53 & \cellcolor{concat15}63.78 & 
  \cellcolor{reno20}71.43 & \cellcolor{reno20}70.89 & 
  \cellcolor{concat5}52.83  & \cellcolor{concat5}51.93 & 
  \cellcolor{reno15}63.42 & \cellcolor{reno15}62.90 \\
H+U   & 
  \cellcolor{concat15}62.57 & \cellcolor{concat15}61.71 & 
  \cellcolor{reno15}66.29 & \cellcolor{reno15}65.72 & 
  \cellcolor{concat15}61.54 & \cellcolor{concat15}60.78 & 
  \cellcolor{reno15}68.21 & \cellcolor{reno15}67.45 & 
  \cellcolor{concat5}54.50  & \cellcolor{concat5}53.02 & 
  \cellcolor{reno10}56.06 & \cellcolor{reno10}55.22 \\
H+W2  & 
  \cellcolor{concat15}64.72 & \cellcolor{concat15}63.90 & 
  \cellcolor{reno15}67.53 & \cellcolor{reno15}66.01 & 
  \cellcolor{concat15}63.65 & \cellcolor{concat15}62.73 & 
  \cellcolor{reno15}69.78 & \cellcolor{reno15}68.65 & 
  \cellcolor{concat5}52.41  & \cellcolor{concat5}51.63 & 
  \cellcolor{reno10}59.71 & \cellcolor{reno10}58.64 \\
H+W   & 
  \cellcolor{concat10}59.72 & \cellcolor{concat10}58.34 & 
  \cellcolor{reno20}63.09 & \cellcolor{reno20}62.21 & 
  \cellcolor{concat15}67.72 & \cellcolor{concat15}66.49 & 
  \cellcolor{reno20}74.51 & \cellcolor{reno20}72.57 & 
  \cellcolor{concat10}55.60 & \cellcolor{concat10}52.71 & 
  \cellcolor{reno20}64.63 & \cellcolor{reno20}62.20 \\
\bottomrule
\end{tabular}
\caption{Evaluation scores are in \% and average of five folds}
\label{tab:2}
\end{table*}

\section{EXPERIMENTS \& RESULTS}

\subsection{Benchmark Dataset}

\noindent 
\textbf{ASVP\_ESD}: \cite{landry2020asvp}: It comprises thousands of high-quality audio recordings labeled with 12 distinct emotions, along with an additional ``breath'' category. These recordings were captured in real-world environments. In our study, we specifically utilize only the non-speech component. The audio samples were sourced from a variety of media, including movies, television shows, YouTube channels, and other online platforms. \par
\noindent \textbf{JNV} \cite{xin2024jnv}:  The dataset comprises 420 high-quality nonverbal vocalization samples, recorded from four native Japanese speakers (two male, two female). The dataset covers six distinct emotions—anger, disgust, fear, happiness, sadness, and surprise—and includes 87 unique phrases. \par
\noindent \textbf{VIVAE} \cite{holz2022variably}: The dataset comprises 1085 high-quality audio samples of non-speech vocalizations, recorded from eleven female speakers in a controlled studio environment. Each sample captures one of six distinct emotional states—achievement/triumph, sexual pleasure, surprise, anger, fear, and physical pain—expressed at four intensity levels: low, moderate, strong, and peak. \par

\noindent \textbf{Training Details}: The models were trained using Adam optimizer with learning rate of 1e-3, a batch size of 32, and 20 epochs. It uses cross-entropy as the classification loss. We use dropout and early stopping for preventing overfitting. For experiments with \textbf{\texttt{RENO}}, we fix \( \beta = 2 \), \( \delta = 0.2 \), and \( \lambda = 0.4 \) for all experiments, as preliminary exploration indicated that these values yielded the best results. We follow five fold cross validation for training our models where four folds are used for training and one fold for testing.  \par
 
\subsection{Results and Discussion}

\begin{figure}[hbt!]
    \centering
    \begin{minipage}{0.22\textwidth}
        \centering
        \includegraphics[width=\textwidth]{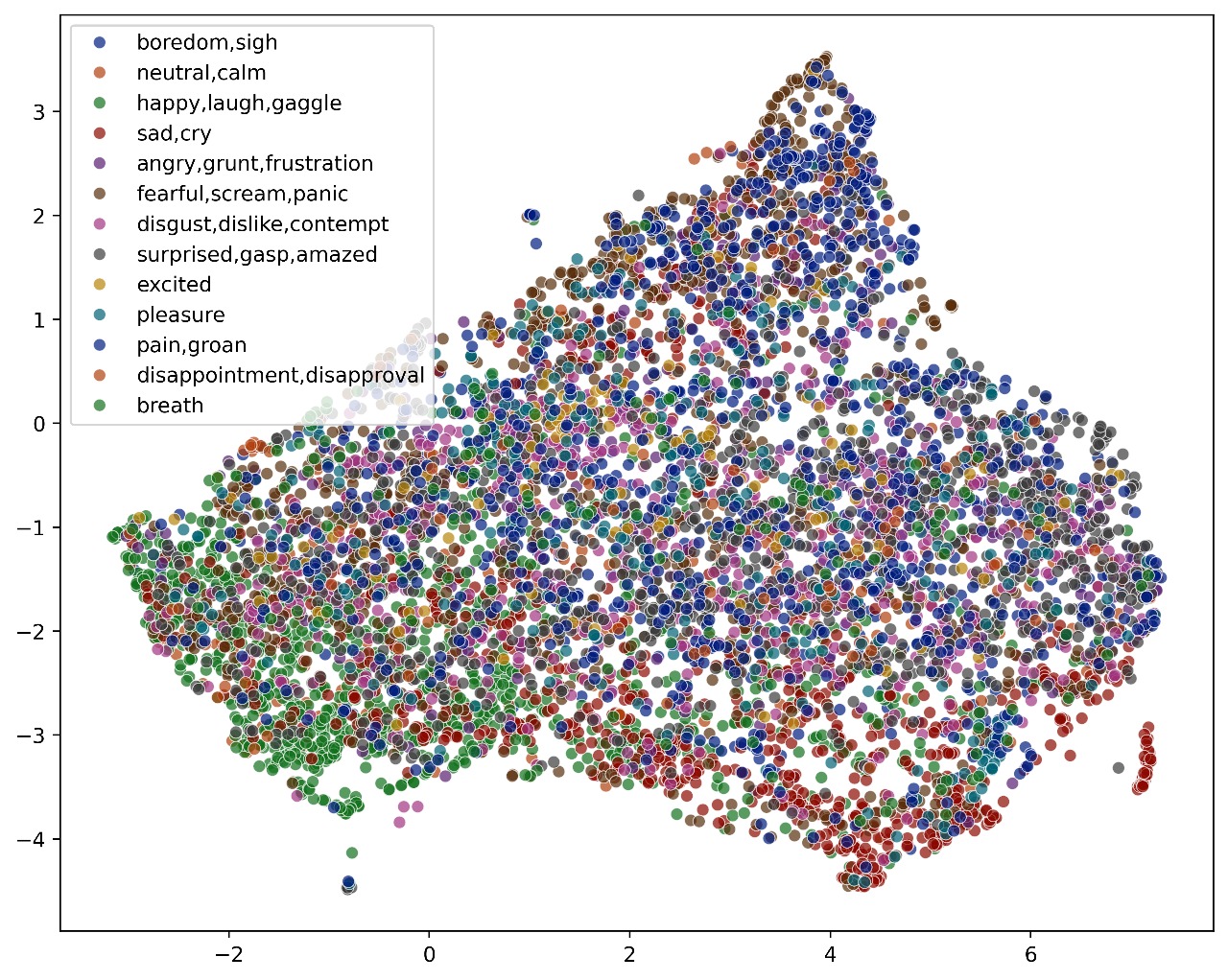} 
        \caption*{(a) Wav2vec2}
    \end{minipage}\hfill
    \begin{minipage}{0.22\textwidth}
        \centering
        \includegraphics[width=\textwidth]{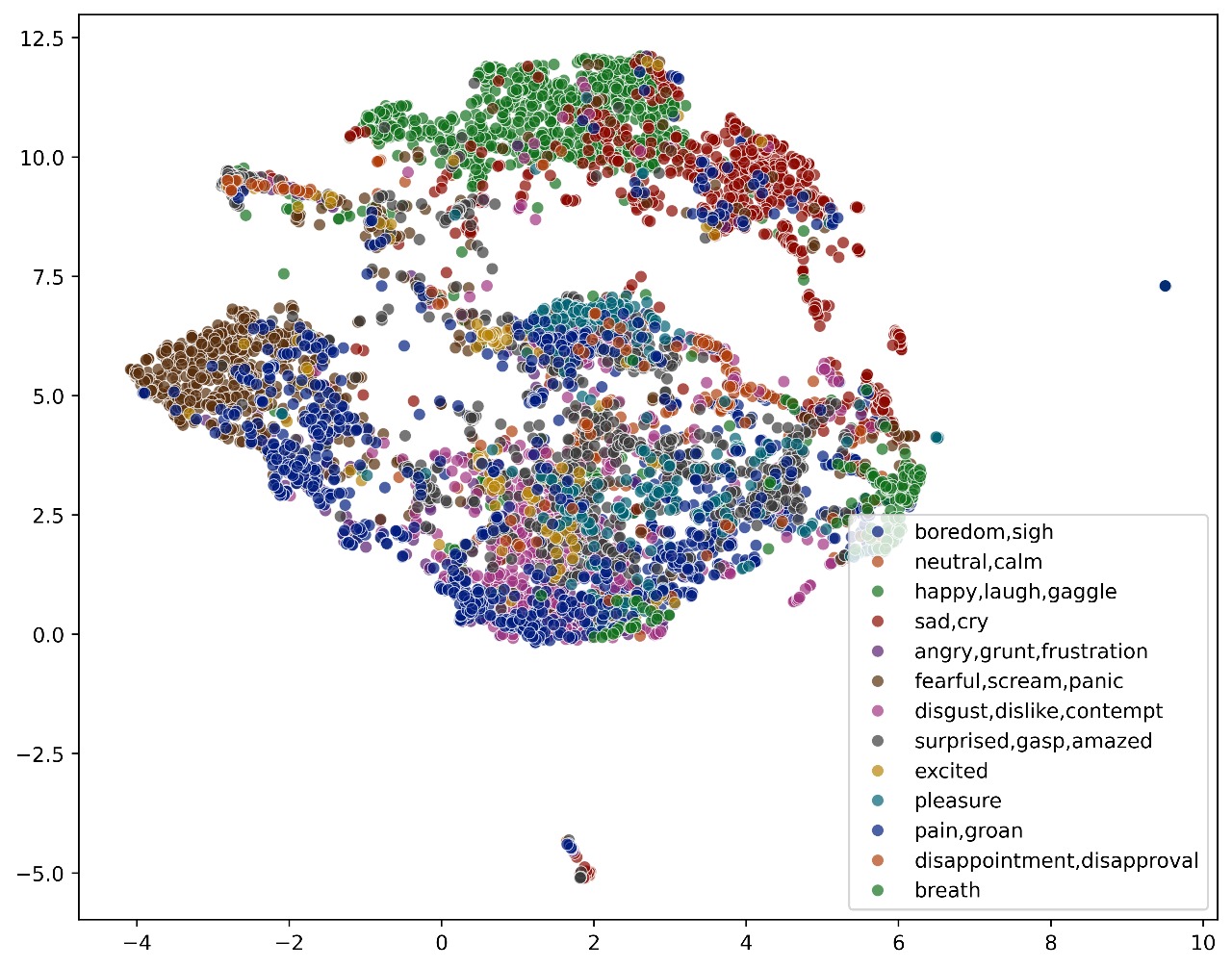} 
        \caption*{(b) Audio-mamba (Base)}
    \end{minipage}\\[10pt] 

    \caption{t-SNE plots for ASVP-ESD} 
    \label{fig:tsne}
\end{figure}

The evaluation scores for downstream models trained on SOTA MAFMs and AAFMs are given in Table~\ref{tab-1}. Our results reveals that MAFMs consistently outperform AAFMs for NVER achieving the highest accuracy and F1-score across all datasets. The superior performance of MAFMs is attributed to their structured state-space modeling, which efficiently captures long-range dependencies and provides more stable and context-aware emotional representations and thus proving our hypothesis. Among the MAFMs, the Audio-mamba (Base) showed the best performance and this can be due to its larger size in comparison to small and tiny. One interesting observation is that despite Audio-mamba (Tiny) is of 4.8M, it is able to beat large AAFMs. This further amplifies our hypothesis that MAFMs will be the most effective for NVER. Overall, the CNN models showed better performance than FCN models. Among the AAFMs, we observe mix performance, with some AAFMs performing better in one dataset and some in other dataset. We also plot the t-SNE plots of raw representations of Audio-mamba (Base) and Wav2vec2 in Figure \ref{fig:tsne}. We observe better cluster for Audio-mamba (Base), thus amplifying our results. \par

\begin{figure}[hbt!]
    \centering
    \begin{subfigure}[b]{0.22\textwidth} 
        \centering
        \includegraphics[width=\linewidth]{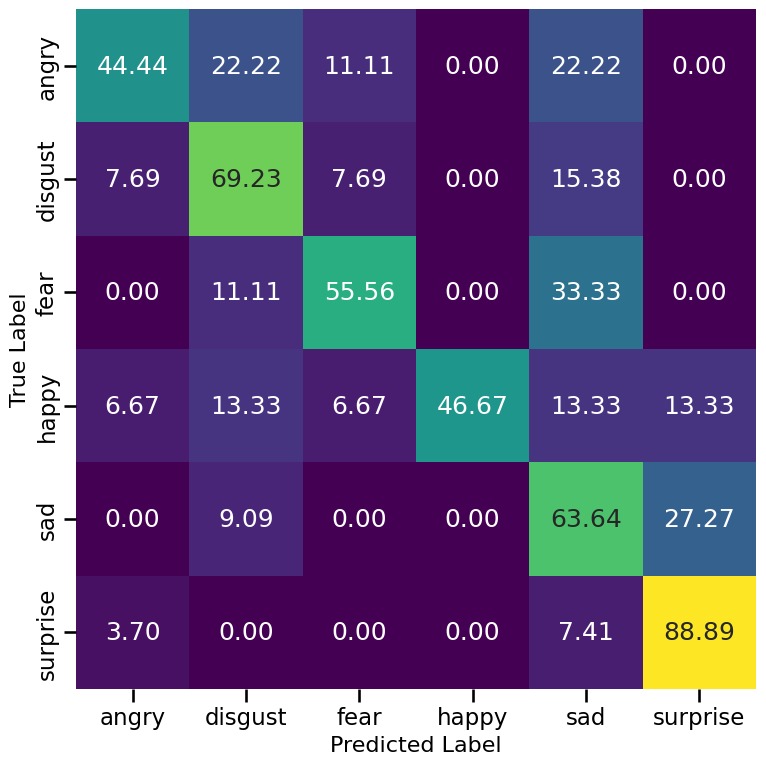}
        \caption{}
    \end{subfigure}
    \hfill
    \begin{subfigure}[b]{0.22\textwidth} 
        \centering
        \includegraphics[width=\linewidth]{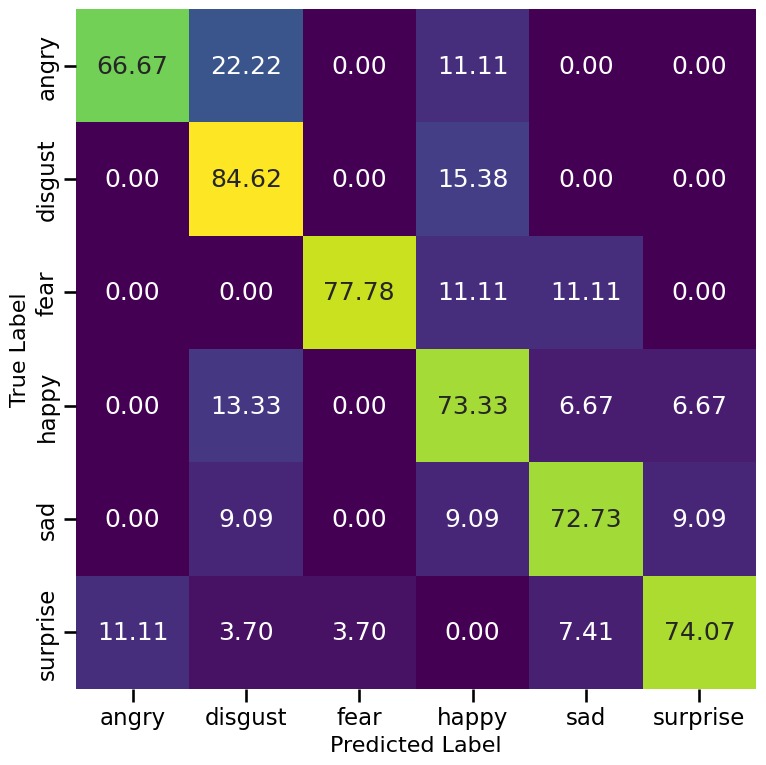}
        \caption{}
        \label{fig:histogram}
    \end{subfigure}
    \caption{Confusion Matrix for JNV dataset; Subfigures (a) Audio-Mamba (Base) with CNN (b) Fusion of Audio-mamba (Tiny) and HuBERT through \textbf{\texttt{RENO}}}
    \label{fig:confusion_matrices}
\end{figure}

Table \ref{tab:2} presents the evaluation scores for different combinations of FMs. We refrain from combining the MAFMs, as it the same model except slight difference in architecture and pre-training data. We use concatenation-based fusion as the baseline fusion technique. We follow the same architectural details as \textbf{\texttt{RENO}} except the renyi rivergence loss and the self-attention blocks. We also keep the training details same as \textbf{\texttt{RENO}}. Our results shows that fusion with \textbf{\texttt{RENO}} leads to better performance in comparison to concatenation-based fusion technique as well as individual MAFMs and AAFMs. We also observe that heterogeneous fusion of MAFMs and AAFMs generally leads to improved performance in comparison to homogeneous fusion of AAFMs. This points out towards observable emergence of complementary behavior as both of them have their own unique strengths. With this heterogeneous fusion of MAFMs and AAFMs through \textbf{\texttt{RENO}}, we set the new SOTA for NVER. However, there is no clear winner among which is the best pair for all the NVER datasets considered, as a particular pair shows the topmost in one dataset and some other pair in another dataset. For example, combination of Audio-Mamba (Base) and Unispeech-SAT with \textbf{\texttt{RENO}} is leading in ASVP-ESD but Audio-Mamba (Base) and HuBERT is top in VIVAE. This behavior is most possibly to the dependence on downstream data distribution variability. We plot the confusion matrices of CNN model built on Audio-mamba (Base) and fusion of Audio-mamba (Tiny) with HuBERT through \textbf{\texttt{RENO}} in Figure \ref{fig:confusion_matrices}.


\section{CONCLUSION}
In this work, we establish the potential of MAFMs for NVER, demonstrating their superiority over AAFMs in capturing intrinsic emotional structures through state-space modeling. Unlike AAFMs, which may amplify irrelevant patterns, MAFMs provide stable, context-aware representations, enhancing the recognition of subtle non-verbal emotional cues. Our experiments validate this hypothesis, showing MAFMs outperforming SOTA AAFMs. Additionally, inspired by advances in SER and synthetic speech detection, we explore FM fusion for NVER. To this end, we introduce \texttt{\textbf{RENO}} for effective fusion of FMs. Through the heterogeneous fusion of MAFMs and AAFMs, \texttt{\textbf{RENO}}  achieves the best performance, surpassing individual FMs, fusion baselines, and sets SOTA for NVER. Our study will act as a baseline for future research exploring FMs for NVER. \newline
\vspace{-3.1mm}

\bibliographystyle{IEEEtran}
\bibliography{main}

\begin{thebibliography}{10}
\providecommand{\url}[1]{#1}
\csname url@samestyle\endcsname
\providecommand{\newblock}{\relax}
\providecommand{\bibinfo}[2]{#2}
\providecommand{\BIBentrySTDinterwordspacing}{\spaceskip=0pt\relax}
\providecommand{\BIBentryALTinterwordstretchfactor}{4}
\providecommand{\BIBentryALTinterwordspacing}{\spaceskip=\fontdimen2\font plus
\BIBentryALTinterwordstretchfactor\fontdimen3\font minus \fontdimen4\font\relax}
\providecommand{\BIBforeignlanguage}[2]{{%
\expandafter\ifx\csname l@#1\endcsname\relax
\typeout{** WARNING: IEEEtran.bst: No hyphenation pattern has been}%
\typeout{** loaded for the language `#1'. Using the pattern for}%
\typeout{** the default language instead.}%
\else
\language=\csname l@#1\endcsname
\fi
#2}}
\providecommand{\BIBdecl}{\relax}
\BIBdecl

\bibitem{lin2005speech}
Y.-L. Lin and G.~Wei, ``Speech emotion recognition based on hmm and svm,'' in \emph{2005 international conference on machine learning and cybernetics}, vol.~8.\hskip 1em plus 0.5em minus 0.4em\relax IEEE, 2005, pp. 4898--4901.

\bibitem{kishore2013emotion}
K.~K. Kishore and P.~K. Satish, ``Emotion recognition in speech using mfcc and wavelet features,'' in \emph{2013 3rd IEEE International Advance Computing Conference (IACC)}.\hskip 1em plus 0.5em minus 0.4em\relax IEEE, 2013, pp. 842--847.

\bibitem{liu2018speech}
Z.-T. Liu, M.~Wu, W.-H. Cao, J.-W. Mao, J.-P. Xu, and G.-Z. Tan, ``Speech emotion recognition based on feature selection and extreme learning machine decision tree,'' \emph{Neurocomputing}, vol. 273, pp. 271--280, 2018.

\bibitem{wang2020speech}
J.~Wang, M.~Xue, R.~Culhane, E.~Diao, J.~Ding, and V.~Tarokh, ``Speech emotion recognition with dual-sequence lstm architecture,'' in \emph{ICASSP 2020-2020 IEEE International Conference on Acoustics, Speech and Signal Processing (ICASSP)}.\hskip 1em plus 0.5em minus 0.4em\relax IEEE, 2020, pp. 6474--6478.

\bibitem{pawar2021convolution}
M.~D. Pawar and R.~D. Kokate, ``Convolution neural network based automatic speech emotion recognition using mel-frequency cepstrum coefficients,'' \emph{Multimedia Tools and Applications}, vol.~80, pp. 15\,563--15\,587, 2021.

\bibitem{lian2021ctnet}
Z.~Lian, B.~Liu, and J.~Tao, ``Ctnet: Conversational transformer network for emotion recognition,'' \emph{IEEE/ACM Transactions on Audio, Speech, and Language Processing}, vol.~29, pp. 985--1000, 2021.

\bibitem{pepino21_interspeech}
L.~Pepino, P.~E. Riera, and L.~Ferrer, ``Emotion recognition from speech using wav2vec 2.0 embeddings,'' \emph{ArXiv}, vol. abs/2104.03502, pp. 3400--3404, 2021.

\bibitem{phukan2023transforming}
O.~C. Phukan, A.~B. Buduru, and R.~Sharma, ``Transforming the embeddings: A lightweight technique for speech emotion recognition tasks,'' \emph{arXiv preprint arXiv:2305.18640}, 2023.

\bibitem{diatlova2024adapting}
D.~Diatlova, A.~Udalov, V.~Shutov, and E.~Spirin, ``Adapting wavlm for speech emotion recognition,'' in \emph{Proc. odyssey 2024}, 2024, pp. 303--308.

\bibitem{yadav24_interspeech}
S.~Yadav and Z.-H. Tan, ``Audio mamba: Selective state spaces for self-supervised audio representations,'' in \emph{Interspeech 2024}, 2024, pp. 552--556.

\bibitem{hsu2021speech}
J.-H. Hsu, M.-H. Su, C.-H. Wu, and Y.-H. Chen, ``Speech emotion recognition considering nonverbal vocalization in affective conversations,'' \emph{IEEE/ACM Transactions on Audio, Speech, and Language Processing}, vol.~29, pp. 1675--1686, 2021.

\bibitem{xin2024jvnv}
D.~Xin, J.~Jiang, S.~Takamichi, Y.~Saito, A.~Aizawa, and H.~Saruwatari, ``Jvnv: A corpus of japanese emotional speech with verbal content and nonverbal expressions,'' \emph{IEEE Access}, vol.~12, pp. 19\,752--19\,764, 2024.

\bibitem{tzirakis2023large}
P.~Tzirakis, A.~Baird, J.~Brooks, C.~Gagne, L.~Kim, M.~Opara, C.~Gregory, J.~Metrick, G.~Boseck, V.~Tiruvadi \emph{et~al.}, ``Large-scale nonverbal vocalization detection using transformers,'' in \emph{ICASSP 2023-2023 IEEE International Conference on Acoustics, Speech and Signal Processing (ICASSP)}.\hskip 1em plus 0.5em minus 0.4em\relax IEEE, 2023, pp. 1--5.

\bibitem{10317508}
Y.~Wu, P.~Yue, C.~Cheng, and T.~Li, ``Investigation of ensemble of self-supervised models for speech emotion recognition,'' in \emph{2023 Asia Pacific Signal and Information Processing Association Annual Summit and Conference (APSIPA ASC)}, 2023, pp. 988--995.

\bibitem{chetia-phukan-etal-2024-heterogeneity}
\BIBentryALTinterwordspacing
O.~Chetia~Phukan, G.~Kashyap, A.~B. Buduru, and R.~Sharma, ``Heterogeneity over homogeneity: Investigating multilingual speech pre-trained models for detecting audio deepfake,'' in \emph{Findings of the Association for Computational Linguistics: NAACL 2024}, K.~Duh, H.~Gomez, and S.~Bethard, Eds.\hskip 1em plus 0.5em minus 0.4em\relax Mexico City, Mexico: Association for Computational Linguistics, Jun. 2024, pp. 2496--2506. [Online]. Available: \url{https://aclanthology.org/2024.findings-naacl.160/}
\BIBentrySTDinterwordspacing

\bibitem{chen2022wavlm}
S.~Chen, C.~Wang, Z.~Chen, Y.~Wu, S.~Liu, Z.~Chen, J.~Li, N.~Kanda, T.~Yoshioka, X.~Xiao \emph{et~al.}, ``Wavlm: Large-scale self-supervised pre-training for full stack speech processing,'' \emph{IEEE Journal of Selected Topics in Signal Processing}, vol.~16, no.~6, pp. 1505--1518, 2022.

\bibitem{chen2022unispeech}
S.~Chen, Y.~Wu, C.~Wang, Z.~Chen, Z.~Chen, S.~Liu, J.~Wu, Y.~Qian, F.~Wei, J.~Li, and X.~Yu, ``Unispeech-sat: Universal speech representation learning with speaker aware pre-training,'' \emph{ICASSP 2022 - 2022 IEEE International Conference on Acoustics, Speech and Signal Processing (ICASSP)}, pp. 6152--6156, 2021.

\bibitem{baevski2020wav2vec}
A.~Baevski, Y.~Zhou, A.~Mohamed, and M.~Auli, ``wav2vec 2.0: A framework for self-supervised learning of speech representations,'' \emph{Advances in neural information processing systems}, vol.~33, pp. 12\,449--12\,460, 2020.

\bibitem{hsu2021hubert}
W.-N. Hsu, B.~Bolte, Y.-H.~H. Tsai, K.~Lakhotia, R.~Salakhutdinov, and A.~Mohamed, ``Hubert: Self-supervised speech representation learning by masked prediction of hidden units,'' \emph{IEEE/ACM transactions on audio, speech, and language processing}, vol.~29, pp. 3451--3460, 2021.

\bibitem{van2014renyi}
T.~Van~Erven and P.~Harremos, ``R{\'e}nyi divergence and kullback-leibler divergence,'' \emph{IEEE Transactions on Information Theory}, vol.~60, no.~7, pp. 3797--3820, 2014.

\bibitem{landry2020asvp}
D.~Landry, Q.~He, H.~Yan, and Y.~Li, ``Asvp-esd: A dataset and its benchmark for emotion recognition using both speech and non-speech utterances,'' \emph{Global Scientific Journals}, vol.~8, pp. 1793--1798, 2020.

\bibitem{xin2024jnv}
D.~Xin, S.~Takamichi, and H.~Saruwatari, ``Jnv corpus: A corpus of japanese nonverbal vocalizations with diverse phrases and emotions,'' \emph{Speech Commun.}, vol. 156, p. 103004, 2023.

\bibitem{holz2022variably}
N.~Holz, P.~Larrouy-Maestri, and D.~Poeppel, ``The variably intense vocalizations of affect and emotion (vivae) corpus prompts new perspective on nonspeech perception.'' \emph{Emotion}, vol.~22, no.~1, p. 213, 2022.

\end{thebibliography}

\end{document}